# Future HEP Accelerators: The US Perspective

Pushpalatha BHAT, Vladimir SHILTSEV

*Fermilab [1], PO Box 500, Batavia IL, 60510, USA*

Accelerator technology has advanced tremendously since the introduction of accelerators in the 1930's, and particle accelerators have become indispensable instruments in high energy physics (HEP) research to probe Nature at smaller and smaller distances. At present, accelerator facilities can be classified into "Energy Frontier" colliders that enable direct discoveries and studies of high mass scale particles and "Intensity Frontier" accelerators for exploration of extremely rare processes, usually at relatively low energies. The near term strategies of the global energy frontier particle physics community are centered on fully exploiting the physics potential of the Large Hadron Collider (LHC) at CERN through its high-luminosity upgrade (HL-LHC), while the intensity frontier HEP research is focused on studies of neutrinos at the MW-scale beam power accelerator facilities, such as Fermilab's Main Injector with the planned PIP-II SRF linac project. A number of next generation accelerator facilities have been proposed and are currently under consideration for the medium- and long-term future programs of accelerator-based HEP research. In this paper, we briefly review the post-LHC energy frontier options, both for lepton and hadron colliders in various regions of the world, as well as possible future intensity frontier accelerator facilities.

---

[1] Fermi Research Alliance, LLC operates Fermilab under contract no. DE-AC02-07CH11359 with the U.S. Department of Energy.

# 1 Introduction

Progress in elementary particle physics over the past fifty years has come mostly from research and discoveries at successively more powerful accelerators, particularly, lepton and hadron colliders. New accelerator concepts and breakthrough technologies have frequently provided cost effective ways to get to higher beam energies and luminosities at the colliders. The discovery of the Higgs boson [1] by the ATLAS and CMS collaborations in the maiden run (2010-12, √s = 7, 8 TeV) of the Large Hadron Collider (LHC) completes the Standard Model (SM) of particle physics, the theoretical framework developed in the second half of the twentieth century. All matter and force-carrier particles expected in the SM have now been discovered, a majority of them in experiments at particle accelerators and colliders, and the predictions of the SM confirmed with remarkable precision in hundreds of measurements.

While the properties of the Higgs boson, so far, seem consistent with those expected of a SM Higgs boson, its measured mass of ~125 GeV is also consistent with it being the harbinger of new physics beyond the SM (BSM). Therefore, the discovery has reignited strong interest in the world-wide high energy physics community in future energy-frontier colliders beyond the LHC to study the properties of the Higgs boson with great precision and to access BSM physics. In the near term, however, it is of utmost importance that the physics potential of the LHC is exploited to the fullest possible extent. The LHC has now begun its second run with higher collision energy of √s = 13 TeV. As per the current plan, after delivering about 300 fb$^{-1}$ of collision data to the experiments through 2023 (ten times the integrated luminosity of Run 1), the LHC will undergo a major upgrade (High Luminosity or HL-LHC) to deliver another order of magnitude more luminosity integrated over a ten year period. The global HEP community is vigorously exploring several options for "post-LHC" energy frontier colliders that include both lepton and hadron colliders.

The other major thrust in particle physics is the study of neutrinos and of rare processes such as muon to electron conversion. The discovery of neutrino oscillations which indicates that neutrinos have non-zero mass (requiring BSM physics) and the recent meaurement of a surprisingly large value for the neutrino mixing angle $\theta_{13}$ have propelled a campaign for long-baseline neutrino experiments with powerful new accelerator facilities to study neutrino oscillations, mass hierarchy and CP violation in the neutrino sector. These explorations require intensity-frontier accelerators with multi-MW beam power capabilities.

The Particle Physics Project and Prioritization Panel (P5), a subpanel appointed by the US High Energy Physics Advisory Panel (HEPAP) provided, in May 2014, an updated strategic plan for the US HEP program necessary to realize a twenty-year global vision for the field [2].

## 2  Energy Frontier Colliders

Particle physics goals drove the development of energy frontier colliders over the past five decades, initiating a wide range of innovation in accelerator physics and technology leading to over a 100-fold increase in center of mass energy (for both hadron and lepton colliding facilities) and $10^4$-$10^6$ fold increase in collider luminosity. A total of 29 colliders, shown in Fig.1, have been built during this period.

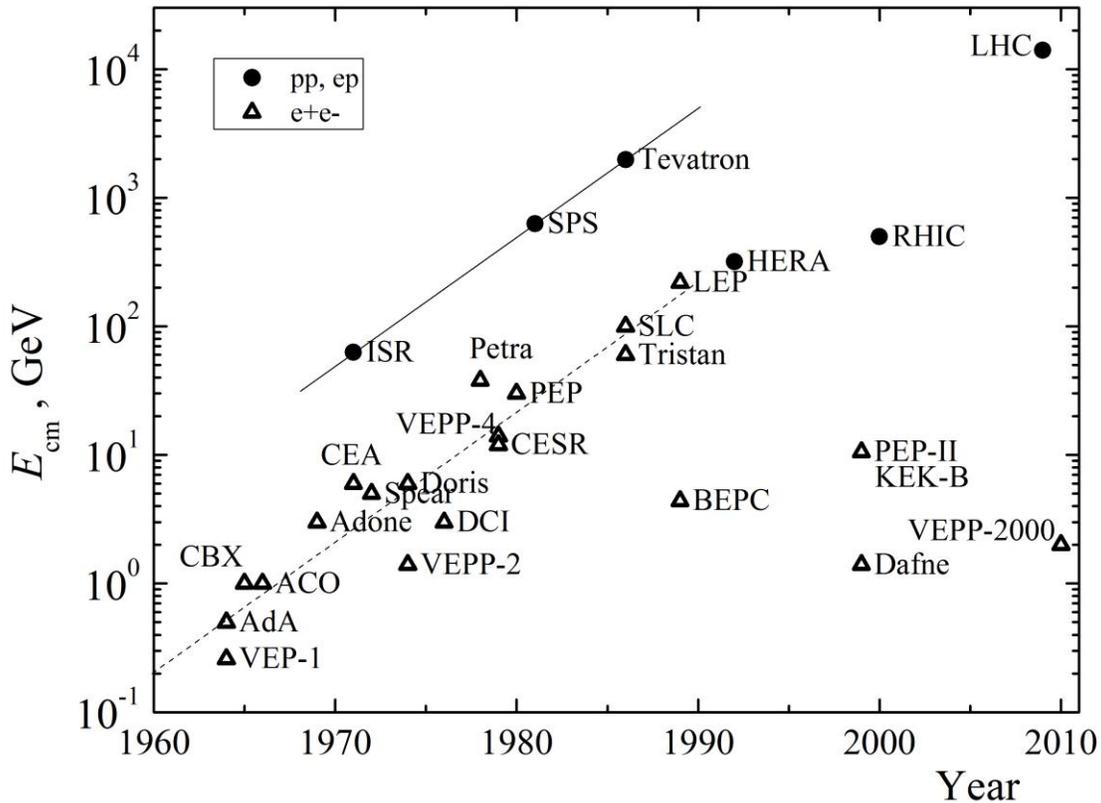

Fig. 1.: Center of mass energies of colliders built over the past five decades as a function of the year of first physics run. [3]

Since the discovery of the Higgs boson, two categories of future colliders [4] have sparked great interest in the HEP community – (1) a "Higgs factory", a lepton collider (most probably an electron-positron collider) with a center of mass (c.m.) energy of 250 GeV and above, and high luminosity, to perform precision studies of the Higgs boson properties, and (2) a proton-proton collider at the "next energy frontier" with collision energies nearly an order of magnitude higher than the LHC energy, say, ~100 TeV, that would help elucidate electroweak symmetry breaking (EWSB) and discover and study particles up to the mass scale of ~50 TeV. The 2014 P5 report underlined the importance of US participation in both categories of these future collider projects. The third category of future colliders of relevance for the study of new physics signals in the TeV range, in a way complementary to the hadron colliders, are multi-TeV lepton colliders.

Table 1: Center of mass energy $E_{cm}$, facility size $L_f$, luminosity $L$ of the future collider projects.

|  | $E_{cm}$, TeV | $L_f$, km | $L$, cm$^{-2}$s$^{-1}$ | Region |
|---|---|---|---|---|
| **CEPC** | 0.25 | 54 | 5·10$^{34}$/IP | China |
| **FCC-ee** | 0.25 | 100 | 5·10$^{34}$/IP | CERN |
| **ILC** | 0.5 | 36 | 2·10$^{34}$ | Japan |
| **CLIC** | 3 | 60 | 5·10$^{34}$ | CERN |
| **µµ-Collider** | 6 | ~20 | 2·10$^{34}$ | US ? |
| **SPPC** | ~50 | 54 | 5·10$^{34}$ | China |
| **FCC-pp/VLHC** | 100 | 100 | 5·10$^{34}$ | CERN/US |

Table 1 presents main parameters of several future colliders under discussion. They are discussed briefly in the following paragraphs.

While there are several options for a lepton collider Higgs factory, shortly after the Higgs discovery, the International Linear Collider (ILC) seemed to be the obvious choice. The ILC concept and design had been well studied for over a decade. A Global Design Effort (GDE) hosted at Fermilab, which was set up in 2008, concluded its work in 2012, after extensive globally coordinated studies and R&D, delivering technical design reports (TDRs) for the ILC accelerator and detectors in June 2013 [5]. The ILC, as designed, stretching over 31 km, with a center of mass energy tunable between 200-500 GeV, and upgradeable to 1 TeV, with luminosity > $10^{34}$cm$^{-2}$s$^{-1}$ at 500 GeV.

With the physics case for an ILC strengthened by the discovery of the Higgs boson, the Japanese HEP community expressed strong interest to host the ILC. Over the past two years, substantial technical progress has been made [6]. Fermilab's superconducting RF (SRF) R&D program has exceeded the ILC accelerating gradient specification of 31.5 MV/m. While the technical challenges seem tractable and can be overcome, funding its construction as an international project appears to be a bigger challenge. The proponents in different world regions are working with government agencies towards an official decision on the project.

Since the mass of the Higgs boson is now known to be ~125 GeV, the alternate and a more lucrative option for an $e+e-$ Higgs factory is to build a circular collider in a large enough circular tunnel (50-100 km circumference) to reach a c.m. energy of up to 250-400 GeV, and at a later time use the tunnel to host the future pp collider that can get to the 70-100 TeV collision energy. This concept is very attractive because of the LEP-LHC experience and success. There is great interest in this concept and ongoing activity with varying degrees of intensity on the design of such circular colliders in Europe (CERN), China and in the U.S.

The Future Circular Colliders (FCC) project [7] activity at CERN follows the European particle physics strategy recommendation and desire, that is, "to propose an ambitious post-LHC project at CERN in a global context". The three options being considered are colliders in a ~100 km tunnel – (1) FCC-ee: an *e+e-* collider with energy from 240-350 GeV, (2) FCC-hh: ~100 TeV pp collider, and (3) FCC-eh: an ep collider. The FCC study is expected to produce a Conceptual Design Report (CDR) and cost estimates for all three options by 2018, in time for the next update of the European strategy for particle physics.

The energy frontier colliders proposed in China [8] are envisioned to be housed in a 50 km ring (although, lately, larger tunnel size is being discussed). The Circular Electron Positron Collider (CEPC) would provide a 90-250 GeV c.m. energy, serving as Z boson and Higgs factory. The Super Proton Proton Collider (SPPC) in the same ring will provide 50 TeV collision energy with 12 T magnets, ~70 TeV if 20 T magnets can be used. China is pursuing a very aggressive timeline, expecting to complete the TDR for CEPC in 2020, starting construction in 2021 and completing construction by 2027. Physics runs with CEPC are anticipated from 2028-35. For SPPC, the design timeline is 2020-30, with a seven-year construction timeline and machine available for physics by 2043. China is expected to undertake these projects as national projects with international cooperation on technology and expertise.

There is revived interest in the US community for hosting a Very Large Hadron Collider (VLHC) in a 100 km ring based at Fermilab [4]. The U.S. had developed a conceptual design report in 2001 for a 233 km VLHC [9], 40 TeV pp collider in the first phase using 2 T transmission line magnets and reaching 200 TeV as a second phase upgrade using 10-12 T superconducting (SC) magnets. The latest preliminary study [4] envisions the use of 16 T SC magnets in a 100 km ring. The Snowmass Accelerator Capabilities report [10] and the 2014 P5 report both support US efforts to strongly participate in the global plans for a 100 km pp collider and be prepared to host it in the US if such opportunity arises. As in the case with FCC-ee/hh at CERN and CEPC/SPPC in China, the aforesaid study also proposes an *e+e-* Higgs factory (VLEP) as the first phase option in the VLHC ring, if the ILC is not realized in the meantime.

All three serious candidates for the Higgs factory (ILC, CEPC, FCC-ee) are based on well developed (whether normal conducting or superconducting) technologies for RF and magnets. Therefore, their feasibility in terms of collision energy is not in doubt. However, the required luminosity performance with $L \sim (2-5) \cdot 10^{34}$ cm$^{-2}$s$^{-1}$ per IP at these colliders is not fully guaranteed due to a number of challenges such as the huge facility power consumption (in the range of 300-500 MW), thermal load in the cold SC RF cavities due to higher-order-mode (HOM heating) and beamstrahlung-limited dynamic aperture for circular *e+e-* colliders CEPC and FCC-ee, and the beam emittance issues in the main linacs and positron production for the ILC. Continuing R&D on these items is essential.

Fig. 2 shows the luminosity performance expected for proposed $e+e-$ collider facilities. The circular $e+e-$ colliders provide substantially higher luminosity at energies below 500 GeV compared to linear colliders and provide multiple interaction points to allow multiple experiments.

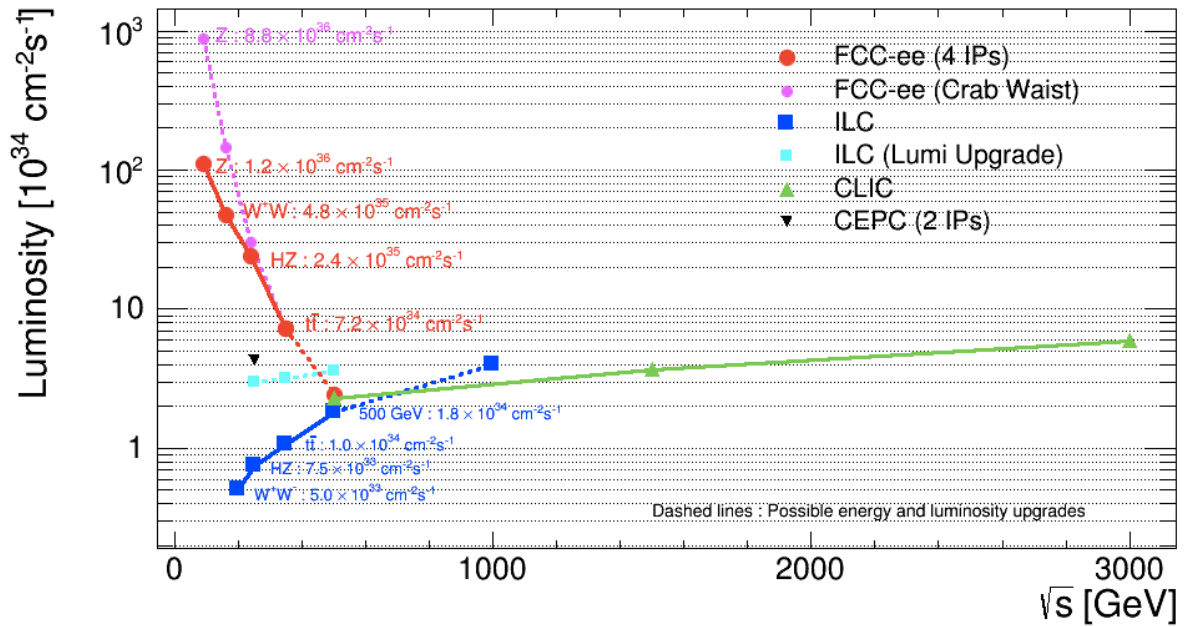

Fig. 2: Expected collider luminosity as a function of collision energy for the proposed $e+e-$ colliders. (From https://tlep.web.cern.ch/)

A major issue for these machines is the ~$10B price tag. Serious efforts need to be undertaken to improve the cost performance [11, 12].

In case of proton-proton supercolliders, such as FCC-hh, VLHC and SPPC, to be declared feasible, several challenges need to be overcome. Achieving the proposed collision energies would require development of ~16T SC magnets which in turn requires vigorous development of $Nb_3Sn$ superconductor technology or of advanced hybrid magnets. The luminosity target of above $5 \cdot 10^{34}$ $cm^{-2}s^{-1}$ is not achievable until critical issues of the synchrotron radiation heat load in the cold magnets, machine protection, ground motion and many others are addressed [13]. The biggest challenge for embarking on building such large machines with 60 to 100 km circumference is their cost. According to a phenomenological cost model [11,12], a 100 km circular collider facility with ~400 MW of site power comsumption and based on today's SC magnets would cost well over $30B, the biggest share of the cost being for the magnets. One of

the primary goals of the long-term R&D program, therefore, should be the development of ~16T (and higher field) SC dipole magnets with cost per TeV (or Tesla-meter) reduced significantly (by a factor 3-5) relative to those of the LHC magnets. (An HE-LHC with 16 T magnets in the LHC tunnel can provide collision energy ~26 TeV.) Another critical issue, in planning these future frontier colliders, is that of training new generations of accelerator physicists and technologists, given the decades-long timescale involved in these projects for design, construction, commissioning and operation.

Of the multi-TeV lepton collider options, the CLIC $e+e-$ collider [14,15] can provide an energy reach of 3 TeV, whereas a muon collider [16], if feasible, can reach well above 3 TeV. Feasibility of the CLIC collider based on the novel two-beam acceleration in 12 GHz normal conducting (NC) RF structures has been demonstrated recently in a small scale CLIC Test Facility (CTF3) where average accelerating gradients of 100 MV/m were achieved with acceptable RF cavity breakdown rates [15]. The luminosity goal of CLIC, $L$~2-5·$10^{34}$ cm$^{-2}$s$^{-1}$ is significantly more challenging than that of the ILC, though the design report indicates no major showstoppers. The biggest issues for CLIC are its enormous site power consuption of about 600 MW and the anticipated cost. Even a six-times smaller version, a 0.5 TeV $e+e-$ collider based on the CLIC technology, has been found to be quite expensive estimated at ~8B Swiss Francs and 14,100-15,700 FTE-years of labor [15]. A muon collider, which is based on the established technology of SC magnets and SRF that can provide energy reach of up to 3-6 TeV, is expected to be relatively much more cost-effective and affordable [11]. Unfortunately, at present, the performance of the muon collider can be assured only at the level two to three orders of magnitude below the design luminosity goal of 2·$10^{34}$ cm$^{-2}$s$^{-1}$ and the performance feasibility requires convincing demonstration of the 6-D ionization cooling of muons. The MICE experiment at RAL is expected to provide the first experimental evidence of muon cooling by 2018.

Finally, the "far future" energy frontier colliders beyond the ones discussed above will require several technological breakthroughs and advanced acceleration techniques. Just as the circular $e+e-$ collider energies cannot be extended beyond the Higgs factory range (~0.25 – 0.4 TeV) due to synchrotron radiation issue, circular proton-proton colliders beyond 100 TeV-scale will become impractical for the same reason. Even an $e+e-$ linear collider becomes impractical above ~3 TeV due to beamstrahlung (radiation due to interaction at the IPs) and at ~10 TeV due to radiation in the focusing channel. This leaves only $\mu+\mu-$ or $pp$ options for colliders beyond these energy regimes. If we were to envision a compact machine, ~10 km in length then we would need an accelerator technolgy that can provide an average gradient of >30 GeV/m (compare with $E/L_f$~ 0.5 GeV/m in the LHC). There is only one such option known now: dense plasma as in, e.g., crystals. This technique excludes protons because of nuclear interactions and leaves us with muons as the particles of choice [3].

There has been much hope and anticipation of plasma-based accelerators – beam-driven or laser-driven wakefields that provide hundred to thousand-fold accelerating gradients than the current state of the art – one day enabling us to build more affordable and compact accelerators/colliders. So far, there are only proof-of-principle experiments for a plasma-wakefield acceleration concept, and several serious challenges are to be overcome to conceptualize such a collider [10].

## 3  Intensity Frontier Accelerators

As recommended in the strategic plan in the 2014 P5 report, the near-term US program of HEP research at the Intensity Frontier continuing throughout this decade and next includes flagship neutrino oscillation physics program and a muon program focused on precision studies of rare processes at Fermilab. It requires: a) doubling the beam power capability of the Booster; b) doubling the beam power capability of the Main Injector; and c) building the muon campus infrastructure and capability based on the 8 GeV proton source. The long-term needs of the Intensity Frontier community are expected to be based on the following experiments: a) long-baseline neutrino experiments to unravel the neutrino sector – study neutrino oscillations, neutrino mass hierarchy and CP-violation in the neutrino sector, and b) precision measurements of rare processes primarily with muons to probe mass-scales beyond that accessible at the LHC.

Fermilab has a strong on-going neutrino physics program using neutrinos produced with the Main Injector beam – MINOS, NOvA and MINERvA, and with the Booster beams – MiniBooNE, SciBooNE (both completed) and MicroBooNE, as well as experiments focused on R&D for future experiments. The proton beam power from the Main Injector on the NUMI target that provides neutrinos for the long baseline neutrino oscillation experiments MINOS and NOvA, has reached over 500 kW. A suite of short baseline neutrino experiments is being prepared. The flagship international neutrino oscillation program – the Long Baseline Neutrino Facility (LBNF) with a Deep Underground Neutrino Experiment (DUNE) at the Homestake mine in South Dakota, 1300 km from Fermilab, is taking shape. CERN has had a rich program of neutrino physics as well, with over 400 kW beam power on target, sending neutrino beams to CNGS with ICARUS and OPERA experiments that took data from 2006 – 2012. CERN is now focussing on a neutrino platform that would enable a rich and very critical R&D program for liquid argon (LAr) TPC program in support of LBNF/DUNE neutrino program based at Fermilab. The other major player in neutrino oscillation physics is the J-PARC program in Japan, currently with a beam power of >300 kW and the upgrades capable of providing 750 kW to 1 MW design power in the coming years. A third generation water Cherenkov detector experiment, Hyper-K (successor of Super-K) is in the plans, to be ready, by the middle of the next decade, for neurtinos from the MW-scale beam from J-PARC.

Fermilab is undertaking a series of upgrades to its accelerator complex, referred to as proton improvement plan (PIP). After completing the ongoing improvements, including the Booster repitition rate upgrade from 7 Hz to 15 Hz, the complex should deliver 700 kW of beam power. Construction of the planned PIP-II SRF 800 MeV proton linac [17] (see Fig. 3) is expected to meet the near-term goals of the intensity frontier program. PIP-II will increase the Booster per pulse beam intensity by 50% and allow delivery of 1.2 MW of the 120 GeV beam power from the Fermilab's Main Injector, with power approaching 1 MW at energies as low as 60 GeV, at the start of DUNE/LBNF operations. It will also support the current 8 GeV program, including the planned suite of short-baseline neutrino experiments, the Mu2e experiment, muon g-2, as well as providing upgrade path for Mu2e and a platform for increasing beam power to multi-MW levels for delivery to DUNE/LBNF.

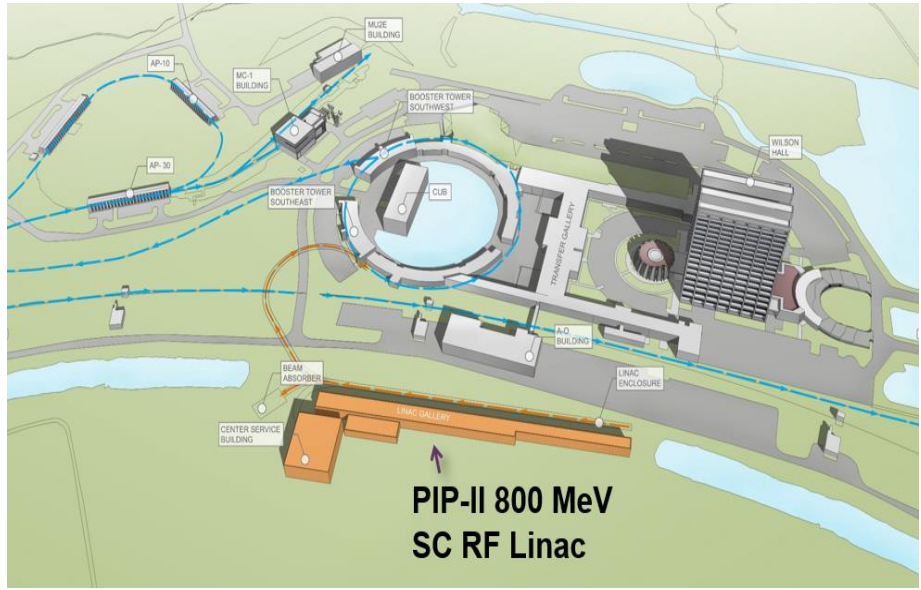

Fig. 3: Artist's rendering of Fermilab accelerator complex with the proposed PIP-II 800 MeV SC linac.

The P5 report recommended a long-term sensitivity goal for the US long-baseline neutrino program as an exposure of 600 kt*MW*yr (the product of detector mass, beam power on target and exposure time). PIP-II offers a platform for the first 100 kt*MW*yr – see Table 2.

Table 2: Neutrino physics program integrated exposure goals and options for achieving them.

|  | **PIP-II** | **Beyond PIP-II (mid-term)** | | |
|---|---|---|---|---|
|  | **1st 10 Years** | **2nd 10 Years** | | |
| Goal ➔ | 100 kt-MW-Year | 500 kt-MW-Year | | |
|  |  | Option 1 | Option 2 | Option 3 |
| Mass | 10 kt | 50 kt | 20 kt | 10 kt |
| Power | 1 MW | 1 MW | 2.5 MW | 5 MW |

The mid-term strategy towards an additional 500 kt*MW*yr beyond PIP-II depends on the technical feasibility of each option shown in Table 2 and the analysis of cost/kiloton of detector versus cost/MW of beam power on target. To make an informed choice, extensive medium-term R&D on the effective control of beam losses in significantly higher current proton machines and on multi-MW targetry are needed [18, 19].

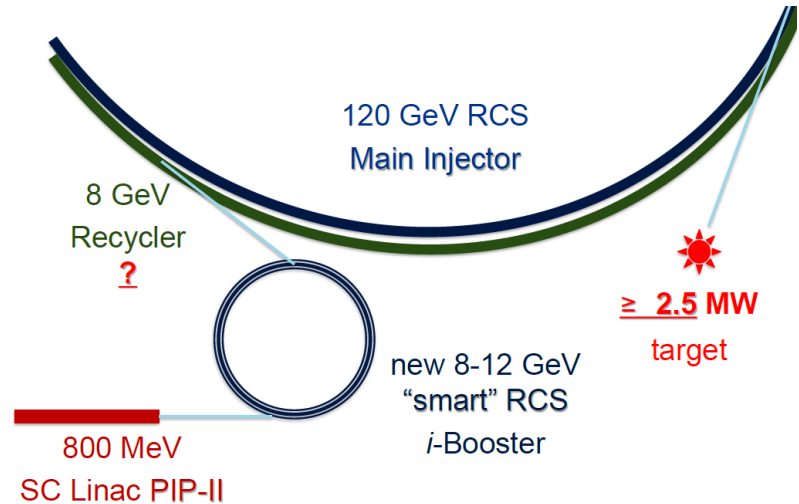

Fig. 4: One of the PIP-III options for the Fermilab accelerator complex upgrade beyond PIP-II assumes replacement of the current Fermilab Booster with a novel high intensity RCS (rapid cycling synchrotron) to enable delivery of ~2.5 MW beam power on the neutrino target for the DUNE experiment.

There are two approaches for the multi-MW proton machine (currently tagged as PIP-III, see Fig. 4) – either a rapid cycling synchrotron or an SRF linac. Achieving the required beam intensities in synchrotrons is only possible if beam losses due to space-charge forces and coherent and incoherent beam instabilities can be significantly reduced. Modern SRF proton linacs can accelerate the required beam but their cost/performance ratio needs to be significantly reduced relative to e.g., the Project X estimates [20]. For both avenues, high-power targetry technology needs to be considerably enhanced to contribute to the feasibility of any multi-MW superbeam facility.

In 2014-15, a sub-panel appointed by HEPAP developed *"A Strategic Plan for Accelerator R&D in the US"* [21] which recommends three R&D activities toward next stage intensity frontier facility:

a) experimental studies of novel techniques to control beam instabilities and particle losses, such as integrable beam optics and space-charge compensation at the Integrable Optics Test Accelerator (IOTA) ring at Fermilab;
b) exploration of the SRF capital and operating cost reductions through transformational R&D on high-Q cavities and innovative materials such as Nb-Cu composites, Nb films and $Nb_3Sn$; cavity performance improvements through novel shapes and field emission elimination;

c) understanding the issues in multi-MW beam targets and developing mitigation techniques, new technologies and new designs.

The IOTA facility at Fermilab [22] is being built as a unique test-bed for transformational R&D towards the next generation high-intensity proton facilities. The accelerator R&D at the IOTA ring with 150 MeV electrons and 70 MeV/c protons, augmented with corresponding modeling and design efforts will lay foundation for novel design concepts, which will allow substantial increase in the proton flux available for HEP research with accelerators to multi-MW beam power levels and also render it cost effective. The goal of the IOTA research program is to carry out experimental studies of techniques to control proton beam instabilities and losses, such as *integrable optics* [23] with non-linear magnets and with electron lenses, and *space-charge compensation* with electron lenses [24] and electron columns [25] at beam intensities and brightness 3-4 times the current operational limits, i.e., at the space-charge parameter approaching or even exceeding $\Delta Q_{SC}$~1.0.

Superconducting RF is the state-of-the-art technology for a majority of future accelerators under consideration due to its unmatched capability to provide up to 100% duty factor, and large apertures to preserve beam quality. The very successful SRF R&D in the past at a number of institutions around the world – Cornell University, DESY in Germany, KEK in Japan, Jefferson Lab and Fermilab – has been predominantly focused on improving gradients, extending from 3 MV/m to >35 MV/m. Recent focus on reducing costs for SRF has led to several major breakthroughs: 1) nitrogen doping for ultra-high $Q$ cavities [26], which opens up more than a factor of two savings in cryogenics capital and operational costs; 2) Nb/Cu composite material and monolithic techniques of cavity manufacturing, which promise more than a factor of two reduction in cavity material and manufacturing costs with performance comparable to bulk Nb cavities; 3) $Nb_3Sn$ cavities for 4.2K operation.

Mega-watt class target facilities present many technical challenges, including issues with radiation damage, rapid heat removal, high thermal shock response, highly non-linear thermo-mechanical effects, radiation protection, and remote handling [27]. The major goal of the R&D program in this area over the next decade is to carry out simulations of high intensity beam/matter interactions using realistic, irradiated material properties to enable designing of multi-MW neutrino and muon target components and systems, and predicting their lifetimes. This requires concerted effort on evaluating relevant materials over a whole host of extreme conditions and environment.

Beyond the MW- and multi-MW superbeams from PIP-II and PIP-III, further future intensity frontier accelerator projects under discussion are centered on neutrino factories [28]. A neutrnio factory would be an intense source of neutrinos from a stored muon beam, an ideal tool for the study of high precision flavour physics. A realistic scenario for a series of staged facilities with increasing complexity and significant physics potential at each stage has been recently developed [29]. It takes advantage of and leverages the capabilities already planned at Fermilab, especially the strategy for long-term improvement of the accelerator complex with the Proton Improvement Plans (PIP-II, III) and the Long Baseline Neutrino

Facility (LBNF). The staging scenario might be as follows: i) nuSTORM [30]: a short-baseline proto-type neutrino factory with a small muon storage ring, enabling a sensitive search for sterile neutrinos and precision neutrino cross-section measurements which are critical input for precision measurements in the long-baseline experiments; ii) NuMAX (Neutrinos from the Muon Accelerator Complex): a long-baseline 5 GeV neutrino factory providing a well-characterized neutrino source that exceeds the capabilities of conventional superbeams and optimized for the DUNE detector at a distance of 1300 km from Fermilab; iii) NuMAX+: a full-intensity neutrino factory upgraded from NuMAX, (see also IDS-NF [31] ) a powerful source to enable precision CP-violation measurements and exploration of new physics in the neutrino sector. Thanks to the great synergies between a neutrino factory and a muon collider, these facilities are complementary and allow beam capabilities and world-leading experimental facilities spanning physics at both the intensity and energy frontiers.

## 4 Summary

Particle accelerators have played a central role in establishing the standard model of particle physics. With all of the expected SM particles now discovered, the field has reached a major milestone and looks to embark on precision studies of the Higgs boson properties, completing the elucidation of electroweak symmetry breaking and exploring new physics beyond the SM. To enable these pursuits, several new accelerator facilities, both at the energy frontier and the intensity frontier are in various stages of active discussion, design and planning world-wide. At the energy frontier, the $e+e-$ colliders such as the ILC, CEPC, FCC-ee are under rigorous planning (hopefully one of them will be built!), and the pp colldiers beyond the LHC (and HL-LHC) such as the SPPC and FCC-pp are envisioned to follow. Studies and R&D are also underway towards realizing these future colliders but more rigorous efforts are needed in this area. If new resonances at the few-TeV scale are discovered at the LHC in the next few years, then CLIC and muon collider might become extermely relevant.

Fermilab, in the near term, is focused on the intensity frontier, planning to replace low energy injector accelerators in the complex, the linac and the booster, with new high beam power proton accelerators (PIP-II SRF linac, PIP-III RCS or SRF linac), with the goal of extenisvely exploring the neutrino sector, and rare processes with muon beams.

All these accelrator projects face both technical and budget challenges. The US (global) accelerator physics community is pursuing critical R&D in physics of high intensity beams, high power targets, SRF acceleration, SC high field magnets and advanced acceleration techniques. Ultimately, the discussion of options for the future HEP accelerators comes to the question of the right balance between the physics reach of the facilities and their technical and cost feasibility [32,33].

# Acknowledgements

The authors thank Dmitri Denisov, Mark Palmer and Chandra Bhat for useful input.